\documentclass[multphys,vecphys]{svmult}

\usepackage{makeidx}         
\usepackage{graphicx}        
\usepackage{multicol}        
\usepackage[bottom]{footmisc}

\makeindex             

\begin{document}

\title*{Hard X-ray emission from the core of the Perseus cluster
  \index{A426} and the thermal content of the radio bubbles}
\titlerunning{Hard flux and bubble content in Perseus}
\author{J.S. Sanders\inst{1}\and
A.C. Fabian\inst{2}}
\institute{Institute of Astronomy, University of Cambridge, Madingley
  Road, Cambridge, CB3 0HA. UK.
\texttt{jss@ast.cam.ac.uk}
\and Institute of Astronomy, University of Cambridge, Madingley
  Road, Cambridge, CB3 0HA. UK. \texttt{acf@ast.cam.ac.uk}}
%
%
\maketitle

We use a very deep 900~ks \emph{Chandra} X-ray observation of the core
of the Perseus cluster to measure and confirm the hard X-ray emission
detected from a previous analysis.  By fitting a model made up of
multiple temperature components plus a powerlaw or hot thermal
component, we map the spatial distribution of the hard flux. We
confirm there is a strong hard excess within the central regions. The
total luminosity in the 2-10 keV band inside 3~arcmin radius is $\sim
5\times 10^{43}$~erg~s$^{-1}$. As a second project we place limits on
the thermal gas content of the X-ray cavities in the cluster core.
This is done by fitting a model made up of multiple components to
spectra from inside and outside of the bubbles, and looking at the the
difference in strength of a component at a particular temperature.
This approach avoids assumptions about the geometry of the core of the
cluster.  Only up to 50~per~cent of the volume of the cavities can be
filled with thermal gas with a temperature of 50~keV.

\section{Introduction}
We found evidence for an additional hard emission from the core of the
Perseus cluster in a 200~ks \emph{Chandra} observation
\cite{sandersper04,sandersnontherm05}, which we interpreted as
nonthermal X-ray emission due to inverse Compton scattering of cosmic
microwave background or infrared seed photons. We now verify the
detection of the hard flux and characterise it better with a longer
900~ks \emph{Chandra} observation \cite{fabianper06}.

The first \emph{Chandra} observations of the cluster also showed that
the X-ray cavities likely generated by the lobes the central AGN 3C\,84
\index{3C84} do not appear to contain thermal volume filling gas below
11~keV \cite{schmidtper02}.  Analysis of the 200~ks data suggested
that further limits were difficult to obtain due to the non-spherical
nature of the cluster core preventing deprojection analyses
\cite{sandersper04}. To limit the thermal content further we here
examine the deep 900~ks data without geometric assumptions.

\section{Hard X-ray emission}
In our previous analysis of a 200~ks observation of the cluster we
used a simple thermal plus powerlaw model to search for any hard
emission in the centre of the cluster \cite{sandersnontherm05}. The
thermal component accounts for thermal gas, and the powerlaw accounts
for hot thermal or nonthermal components. The disadvantage of this
approach is that if there are multiphase components or if projected
hot gas is significant, this will lead to a false powerlaw signal. In
the previous analysis, we simulated the effect of projected emission,
to account for its effects.  There is, however, an extremely extended
H$\alpha$ nebulosity around the central galaxy NGC\,1275
\index{NGC1275} \cite{minkowskiperfil57, lyndsperfil70}, associated
with $\sim 10^9$~M${}_{\odot}$ of cool X-ray emitting gas
\cite{fabianper06}. It may therefore be important to include
contributions from cool X-ray gas in the spectral modelling, otherwise
the X-ray emission at low temperatures may be detected by the powerlaw
component.

Radio observations \cite{sijbringthesis93} indicate that the radio
photon index of the central regions is steep ($\Gamma \geq 2$). If
inverse Compton is the emission mechanism and the electron
distribution continues without a break to $\gamma \sim 1000$, we would
predict X-ray powerlaw emission with a similar steep photon index. Hot
thermal gas would give flatter photon indices if fitted with a
powerlaw model.

\subsection{Analysis}
We here examine a 900~ks long \emph{Chandra} dataset
\cite{fabianper06} of the core of the Perseus cluster. We used the
Contour Binning algorithm \cite{sanderscontbin06} to select regions
containing a signal to noise ratio greater than 500 in the 0.5 to
7~keV band on the ACIS-S3 CCD. The method takes a smoothed map (here
we smoothed the input X-ray image with a top hat kernel with a radius
to give a signal to noise ratio greater than 60), and creates bins
following the surface brightness. A geometric constraint factor of 2
was used to stop the bins becoming too elongated. The regions contain
approximately 250,000 foreground counts.

We extracted spectra from the foreground event files for each
observation from each of the spatial regions. We also extracted
background spectra from blank sky background files, and additional
``background'' spectra to correct for out-of-time events (these were
generated from the foreground event files where the \textsc{chipy}
coordinate of each event on the CCD had been randomised). The
foreground spectra were added together, their responses averaged, and
their backgrounds combined using our previous prescription
\cite{fabianper06}.  We constructed a model made up of
photoelectrically absorbed \textsc{apec} \cite{smithapec01} thermal
components at 0.5, 1, 2, 3, 4 and 8 keV, accounting for the range of
thermal gas observed in the cluster, plus a powerlaw component.  The
thermal components were all fixed in temperature, with free
normalisations. The metallicities of the components were tied together
and free in the fit. The absorption was allowed to vary from bin to
bin. The powerlaw photon index was frozen at a value of 2, with a free
normalisation. The model was fitted to the spectra from each region
between 0.6 and 8~keV.

\begin{figure}
  \centering
  \includegraphics[width=0.95\columnwidth]{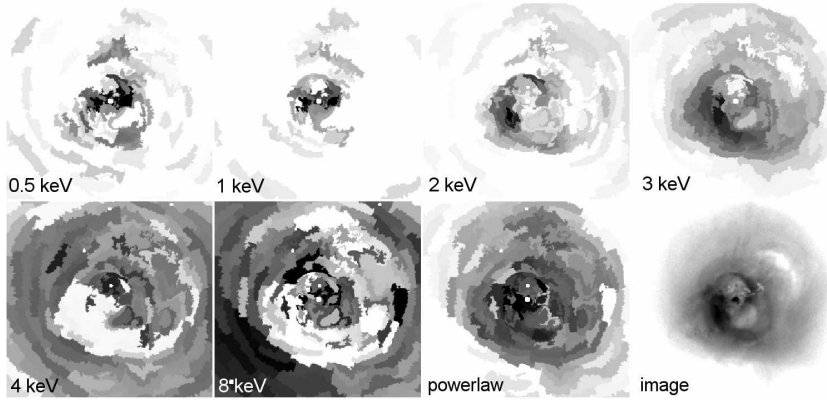}
  \caption{Normalisations of the different temperature components per
    unit area from the spectral fits, plus the normalisation per unit
    area of the powerlaw component with photon index 2. Also shown is
    an X-ray image in the 0.3 to 7~keV band. The images measure
    roughly 4.5 arcmin vertically (100~kpc using if $H_0 =
    70$~km~s$^{-1}$~kpc$^{-1}$).}
  \label{fig:norms}
\end{figure}

Fig.~\ref{fig:norms} shows images of the normalisations per unit area
in the central region for each of the components. As in our previous
analysis \cite{sandersnontherm05}, the normalisation of the powerlaw
component strongly increases in the central regions, indicating it is
not due to a background subtraction problem.

The $\Gamma=2$ fits require that the photoelectric absorption in the
very centre of the cluster, where the powerlaw is strong, is
significantly larger than in the outer regions (an increase of
$N_\mathrm{H}$ of $\sim 2-3 \times 10^{20}$\,cm${}^{-2}$). This is
because the powerlaw component becomes strong at low X-ray energies
releative to the thermal components.

\begin{figure}
  \centering
  \includegraphics[width=0.5\columnwidth]{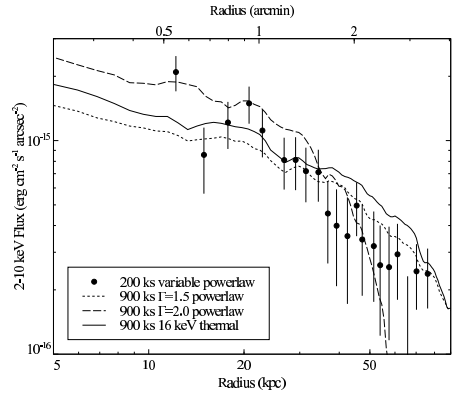}
  \caption{Profile showing the 2-10 keV flux of powerlaw (fixed to
    $\Gamma=2$ or 1.5) or hot thermal (16 keV) component from the fits
    to the new data and the flux from the variable powerlaw fits to
    the old data. Distances assume 0.372 kpc per arcsec.}
  \label{fig:fluxprof}
\end{figure}

If the powerlaw photon index is allowed to be free, steep powerlaws
(if the absorption is free) are preferred over flatter powerlaws.
However, we also attempted to fit models with a flatter powerlaw
photon index ($\Gamma=1.5$) and using a hot 16~keV plasma to replace
the powerlaw component. These fits do not appear to require any excess
absorption if it is free in the fit.  We show profiles of the flux in
the 2-10~keV band for the three different models in
Fig.~\ref{fig:fluxprof}. The plot shows that the models give very
similar fluxes for the central regions of the cluster, indicating the
hard flux measurement is fairly model independent. We also plot a
radial profile measured from the earlier 200~ks observation, using the
simple thermal plus powerlaw model with variable photon index, but
subtracting the modelled contribution from projected gas. The new
results match the old results well.

In the 2-10~keV band, we calculate deabsorbed total fluxes in the
inner 3~arcmin of $5.9 \times 10^{-11}$~erg~cm$^{-2}$~s$^{-1}$ for the
$\Gamma=2$ model, and $6.9 \times 10^{-11}$~erg~cm$^{-2}$~s$^{-1}$ for
the 16~keV thermal model. These fluxes correspond to a 2-10~keV
luminosity of $\sim 5 \times 10^{43}$~erg~s$^{-1}$ if $H_0 =
70$~km~s$^{-1}$~kpc$^{-1}$. This is larger than the luminosity of the
central nucleus in this band \cite{churazovper03}.

\subsection{Conclusions}
We confirm a hard thermal component centred on the core of the Perseus
cluster using a multicomponent spectral model. Its 2-10 keV luminosity
is of the order of $\sim 5 \times 10^{43}$~erg~s$^{-1}$. If the hard
flux is nonthermal in origin with a steep photon index similar to the
radio emission, then additional photoelectric absorption is required
in the centre of the cluster. Flatter nonthermal or hot thermal models
require no additional absorption. In future work we will explore the
different emission mechanisms and their consequences, and examine
other data to limit the possible models.

\section{Limiting hot thermal bubble contents}
To limit the presence of hot thermal gas in the X-ray cavities in the
cluster, we fitted a multicomponent model to spectra extracted from
regions in the holes and other regions at the same radius. We compared
the normalisation of a component of gas a particular temperature
between the two regions.  This differential method avoids geometric
assumptions, only assuming that the cluster is similar at similar
radii. The model consisted of fixed \textsc{apec} components at 0.5,
1, 2, 3, 4 and 5~keV, a powerlaw fixed to $\Gamma=2$ and free
absorption. These components were designed to model the emission from
the cluster plus the hard component above. We also included an
additional component to represent any hot thermal emission in the
bubbles. This component representing the hot gas was stepped in
temperature between 6 and 60 keV. We compared the normalisation per
unit area (for a particular bubble temperature) with another regions
at the same radii, and computed a limit on the difference in
normalisation between the two for each temperature.

The normalisation difference can be converted to a limit on the volume
filling fraction simply. We assume a geometry for the regions
extracted from the bubbles to calculate a volume. The regions on the
sky are 0.12 arcmin in radius, and we assume a cylinder depth 0.42 and
0.6 arcmin for the inner SW and ghost NW bubble, respectively. Using
the volume and the difference in normalisation we can estimate an
upper limit for the density if the thermal gas is volume filling. This
is multiplied by the temperature currently examined to calculate a
volume filling electron pressure upper limit. We took the ratio of
this pressure to the thermal electron pressure of the gas surrounding
the X-ray holes from existing deprojected electron pressure maps
\cite{sandersper04} to calculate the upper limit on the volume filling
fraction.

\begin{figure}
  \centering
  \raisebox{5mm}{
    \includegraphics[width=0.45\columnwidth]{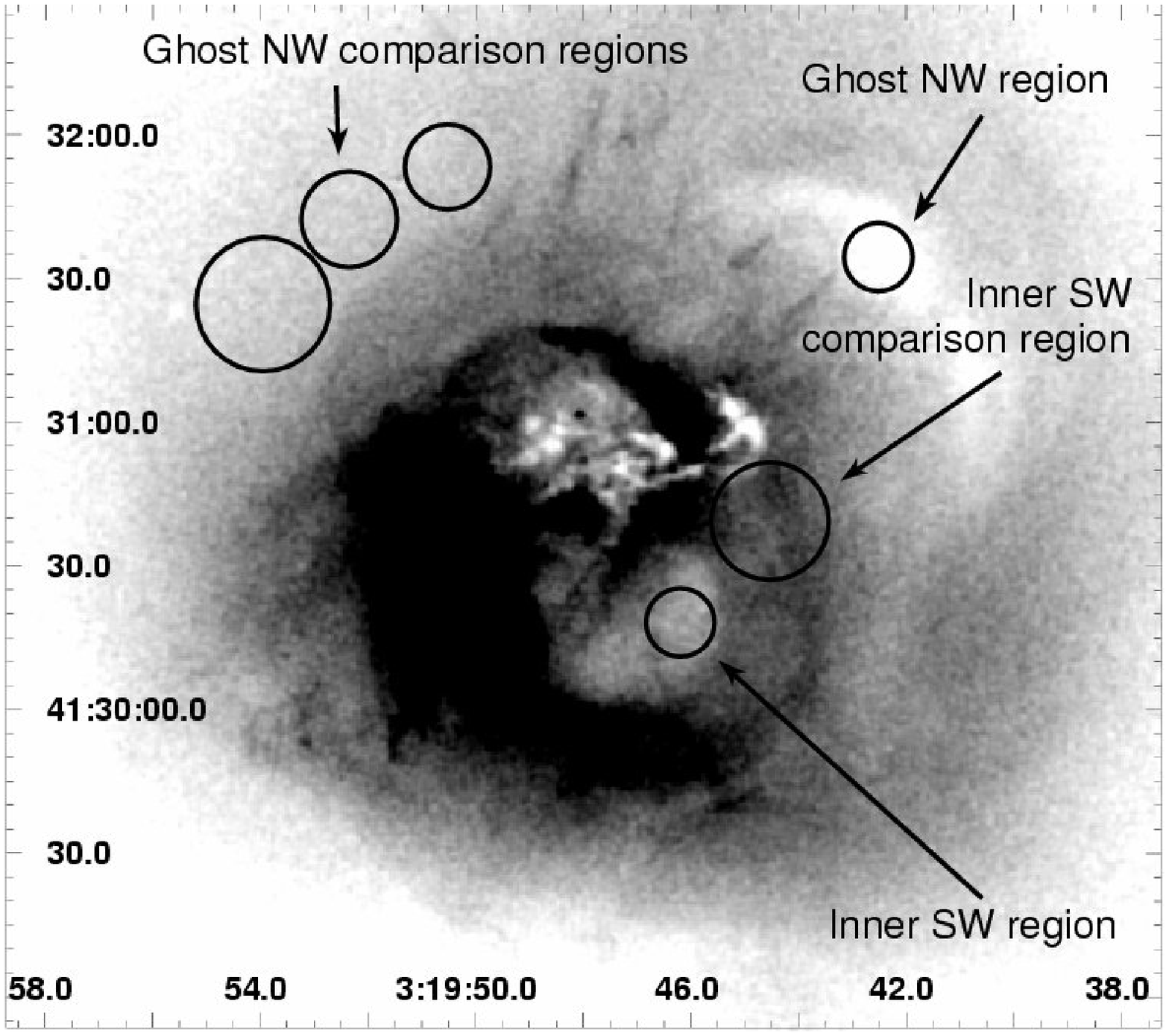}
  }
  \hspace{5mm}
  \includegraphics[width=0.45\columnwidth]{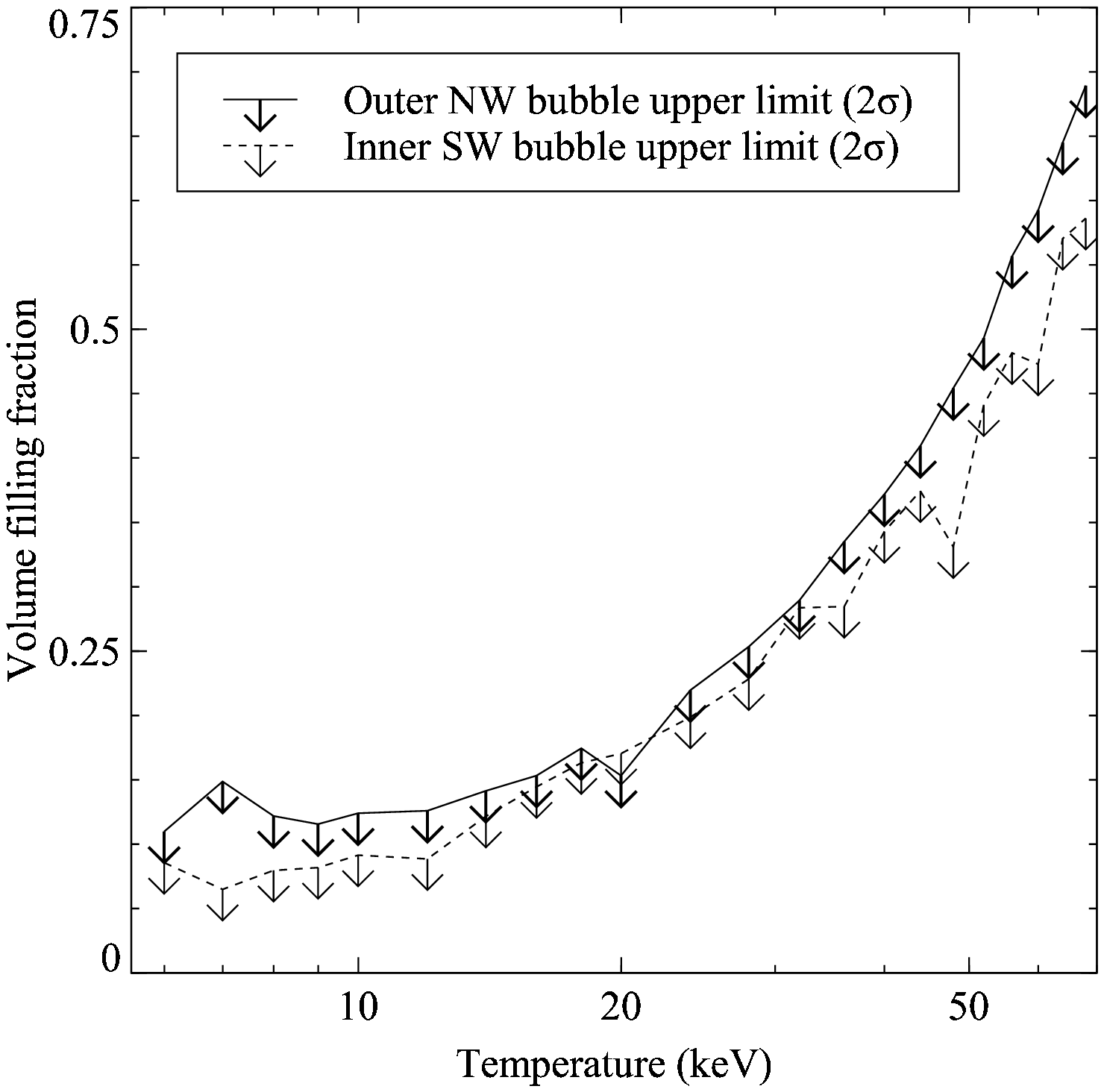}
  \caption{(Left) Bubble and comparison regions in spectral analysis.
    (Right) $2\sigma$ upper limits for the volume filling fraction of
    hot thermal gas within the inner SW radio bubble and outer NW
    ghost cavity.}
  \label{fig:bubblelimits}
\end{figure}

Fig.~\ref{fig:bubblelimits} shows the resulting upper limits as a
function of temperature for the inner SW radio bubble and outer NW
ghost cavity. Note that the limits are not independent. At most half
of the volume of the bubbles is filled with thermal gas with
temperatures less than 50~keV.

\subsection{Conclusions}
We place limits for thermal content of two of the X-ray holes in the
core of the Perseus cluster. By fitting multitemperature models to
regions within the holes and nearby, and comparing the results, we
deduce that at most 50~per~cent of the volume of the bubbles is
occupied by thermal gas below 50~keV.

\printindex

\begin{thebibliography}{99.}
%
%
%


\bibitem{sandersper04} J.S. Sanders, A.C. Fabian, S.W. Allen,
  R.W. Schmidt: MNRAS \textbf{349}, 952 (2004)
\bibitem{sandersnontherm05} J.S. Sanders, A.C. Fabian, R.J.H. Dunn:
  MNRAS \textbf{360}, 133 (2005)
\bibitem{fabianper06} A.C. Fabian, J.S. Sanders, G.B. Taylor,
  S.W. Allen, C.S. Crawford, R.M. Johnstone, K. Iwasawa: MNRAS
  \textbf{366}, 417 (2006)
\bibitem{schmidtper02} R.W. Schmidt, A.C. Fabian, J.S. Sanders: MNRAS
  \textbf{337}, 71 (2002)
\bibitem{minkowskiperfil57} R. Minkowski: Optical investigations of
  radio sources. In: \textit{Proc. IAU Symp. 4}, ed by H.C. Van de
  Hulst (Cambridge Univ. Press, Cambridge 1957), p. 107
\bibitem{lyndsperfil70} R. Lynds: ApJ \textbf{159}, L151 (1970)
\bibitem{sijbringthesis93} D. Sijbring: A Radio Continuum and HI Line
  Study of the Perseus Cluster. PhD Thesis, University of Groningen
  (1993)
\bibitem{sanderscontbin06} J.S. Sanders: MNRAS \textbf{371} 829 (2006)
\bibitem{smithapec01} R.K. Smith, N.S. Brickhouse, D.A. Liedahl,
  J.C. Raymond: ApJ \textbf{556}, L91 (2001)
\bibitem{churazovper03} E. Churazov, W. Forman, C. Jones,
  H. B\"ohringer: ApJ \textbf{590}, 225 (2003)

\end{thebibliography}
\end{document}